\documentclass[english]{article}
\usepackage[utf8]{inputenc}
\usepackage[T1]{fontenc}
\usepackage{babel}
\usepackage{amsmath}
\usepackage{graphicx}
\usepackage{fancyhdr}
\usepackage[margin=4cm]{geometry}

\usepackage{csquotes}
\usepackage[
    backend=biber,
    style=nature,
    url=false,
    doi=true,
    eprint=false
]{biblatex}
\usepackage[affil-it]{authblk}
\usepackage[version=3]{mhchem} % Formula subscripts using \ce{}

\usepackage{color}

\usepackage{float}
\usepackage{graphicx}
\usepackage{subcaption}

\usepackage[hidelinks]{hyperref}
\usepackage{cleveref}
\usepackage{xparse}
\makeatletter
\let\latex@@subsection\subsection
\RenewDocumentCommand{\subsection}{som}{%
  \IfBooleanTF{#1}{%
    \IfValueTF{#2}{%
      \addcontentsline{toc}{subsection}{#2}%
      \latex@@subsection*{#3}\edef\@currentlabelname{#2}%
    }{%
      \latex@@subsection*{#3}\edef\@currentlabelname{#3}%
    }%
  }{%
    \IfValueTF{#2}{%
      \latex@@subsection[#2]{#3}
    }{%
      \latex@@subsection{#3}
    }%
  }%
}
\makeatother

\newcommand{\msection}[1]{{\fontfamily{ptm}\selectfont #1}}
\newcommand{\ARCHIVE}{\msection{ARCHIVE}}
\newcommand{\EXPLORE}{\msection{EXPLORE}}
\newcommand{\LEARN}{\msection{LEARN}}
\newcommand{\DISCOVER}{\msection{DISCOVER}}
\newcommand{\WORK}{\msection{WORK}}

\newcommand{\AIIDALAB}{\msection{AiiDA lab}}
\newcommand{\QM}{\msection{Quantum Mobile}}

\begin{document}

% Authors
\author[1,2,3]{Leopold Talirz\thanks{Corresponding authors: Leopold Talirz (leopold.talirz@gmail.com), Giovanni Pizzi (giovanni.pizzi@epfl.ch), Nicola Marzari (nicola.marzari@epfl.ch)}\thanks{These authors contributed equally.}}
\author[1,2]{Snehal Kumbhar$^\dagger$}
\author[1,2,3]{Elsa Passaro$^\dagger$}
\author[1,2,3]{Aliaksandr V. Yakutovich}
\author[1,2]{Valeria Granata}
\author[1,2]{Fernando Gargiulo}
\author[1,2]{Marco Borelli}
\author[1,2]{Martin Uhrin}
\author[1,2]{Sebastiaan P. Huber}
\author[1,2]{Spyros Zoupanos}
\author[1,2]{Carl S. Adorf}
\author[1,2]{Casper W. Andersen}
\author[1,4]{Ole Schütt}
\author[1,4]{Carlo A. Pignedoli}
\author[1,4]{Daniele Passerone}
\author[1,5,6]{Joost VandeVondele}
\author[1,5,6]{Thomas C. Schulthess}
\author[1,3]{Berend Smit}
\author[1,2]{Giovanni Pizzi$^*$}
\author[1,2]{Nicola Marzari$^*$}

% Affiliations
\affil[1]{National Centre for Computational Design and Discovery
of Novel Materials (MARVEL), \'Ecole Polytechnique F\'ed\'erale de Lausanne, 
CH-1015 Lausanne, Switzerland}
\affil[2]{Theory and Simulation of Materials (THEOS), 
    Facult\'e des Sciences et Techniques de l'Ing\'enieur, 
    \'Ecole Polytechnique F\'ed\'erale de Lausanne,
    CH-1015 Lausanne, Switzerland}
\affil[3]{Laboratory of Molecular Simulation (LSMO),
    Institut des Sciences et Ingenierie Chimiques,
    Valais, \'Ecole Polytechnique F\'ed\'erale de Lausanne,
    CH-1951 Sion, Switzerland}
\affil[4]{nanotech@surfaces laboratory,
    Swiss Federal Laboratories for Materials Science and Technology (Empa),
    CH-8600 D\"ubendorf, Switzerland}
\affil[5]{Swiss National Supercomputing Centre (CSCS),
    CH-6900 Lugano, Switzerland}
\affil[6]{ETH Z\"urich, Switzerland}

% max. 110 characters
\title{Materials Cloud, a platform for open computational science}

\maketitle

% reset to original footnote style
\renewcommand{\thefootnote}{\arabic{footnote}}
\thispagestyle{fancy}
\clearpage

\begin{abstract}

Materials Cloud is a platform designed to enable open and seamless sharing of resources for computational science, driven by applications in materials modelling. 
It hosts
1) archival and dissemination services for raw and curated data, together with their provenance graph,
2) modelling services and virtual machines, 
3) tools for data analytics, and pre-/post-processing,
and 4) educational materials.
Data is citable and archived persistently, providing a comprehensive embodiment of the FAIR principles that extends to computational workflows. 
Materials Cloud leverages the AiiDA framework to record the provenance of entire simulation pipelines (calculations performed, codes used, data generated) in the form of graphs that allow to retrace and reproduce any computed result.
When an AiiDA database is shared on Materials Cloud, peers can browse the interconnected record of simulations, 
download individual files or the full database, and start their research from the results of the original authors. 
The infrastructure is agnostic to the specific simulation codes used and can support diverse applications in computational science that transcend its initial materials domain.

\end{abstract}

%\linenumbers

\section*{Introduction}

Core to the mission of open computational science is the principle that open access to data, software and, eventually, infrastructure leads to scientific results that can be assessed, verified and reproduced.
While this principle has long been at the foundation of science, information technology keeps pushing the limit of what is possible, giving rise to the continuously evolving challenge of translating this principle into practice in a sustainable manner.
Fortunately, funding agencies are increasingly aware of the need to develop comprehensive solutions, including adequate data management plans, \cite{ord-concordat, data-harvest, snf-ord,erc-fdm, amsterdamcall2016} and guidelines are being developed to help ensure that shared resources are easily findable, accessible, interoperable and re-usable (FAIR). \cite{Wilkinson2016}

We believe this challenge calls for \emph{open-science platforms} that let scientists use existing data, submit new content and launch new simulations with minimal requirements on technical expertise.
In this context, it is instructive to look at the field of software engineering, where platforms for sharing source code, such as GitHub (\href{https://github.com}{github.com}), Bitbucket (\href{https://bitbucket.org}{bitbucket.org}), or GitLab (\href{https://gitlab.com}{gitlab.com}) have already revolutionised the industry
-- not only in terms of the volume of source code that is shared publicly, but also in terms of how software developers interact and write code.
These platforms are organised around Git, a software for ``tracking changes in computer files and coordinating work on those files among multiple people''\cite{git-wikipedia}. 
Besides hosting source code repositories, the platforms add a rich web interface for interactive browsing, controlling workflows, and collaboration through social interactions (sharing, commenting, mentioning, etc.).
In our view, open-science platforms can learn from these successful examples, and have the potential to revolutionise the scientific discourse in similar ways.
While these considerations apply to computational science in general,
in the following we focus on the domain of materials.

The field of computational materials science is blessed in that research data in the field is produced in digital form by default,
and many of the necessary computational tools are available free of charge under open-source licenses.
Over the last decade, substantial progress has been made in opening access to some of these resources:
An early example is nanoHUB \cite{Klimeck2008}, which provides access to interactive simulation tools as well as educational materials in the browser.
Platforms have emerged that integrate data repositories with the software frameworks used to compute the data, such as AFLOWlib \cite{Curtarolo2012} (with aflow), the Materials Project \cite{Jain2013} (with pymatgen, custodian, fireworks, atomate), OQMD \cite{Saal2013} (with qmpy), and the Open Materials Database \cite{httk} (with httk).
Finally, there are data repositories, such as NOMAD \cite{Ghiringhelli2017}, that collect and centralise large numbers of individual materials science calculations in one place.

However, the field still faces challenges in the context of open science.
Materials simulations often rely on complex workflows, which, e.g., combine simulations operating at different length- and time-scales or involve cycles of post-processing followed by further simulations.
This calls for a flexible approach to designing such workflows, and to recording their many steps and interconnected results.
Furthermore, screening a class of materials, even for one specific application, may involve running such workflows for thousands of candidate materials or more and require substantial computational power -- the field of computational materials science is among the top consumers of high-performance computing resources around the world \cite{2019k,zotero-24851}.
This makes an efficient and complete record of the workflow execution highly valuable.

In our view, an open-science platform (OSP) should:
\begin{enumerate}
    \item support and adopt open simulation codes and analytics tools; 
    \item provide an open architecture for defining and managing computational workflows;
    \item offer turnkey solutions based on open workflows and curated open datasets
        that are accessible to a diverse user base from computational science, experiments, and industrial R\&D; and
    \item enable FAIR sharing of data and workflows, facilitating reproducibility
        and encouraging extension and/or modification of published resources.
\end{enumerate}
With this vision in mind, we have designed and implemented the Materials Cloud platform (\href{https://www.materialscloud.org}{materialscloud.org}), which we describe in the remainder of this paper.

\section*{Results}

\begin{figure}
\centering
\includegraphics[width=0.85\textwidth]{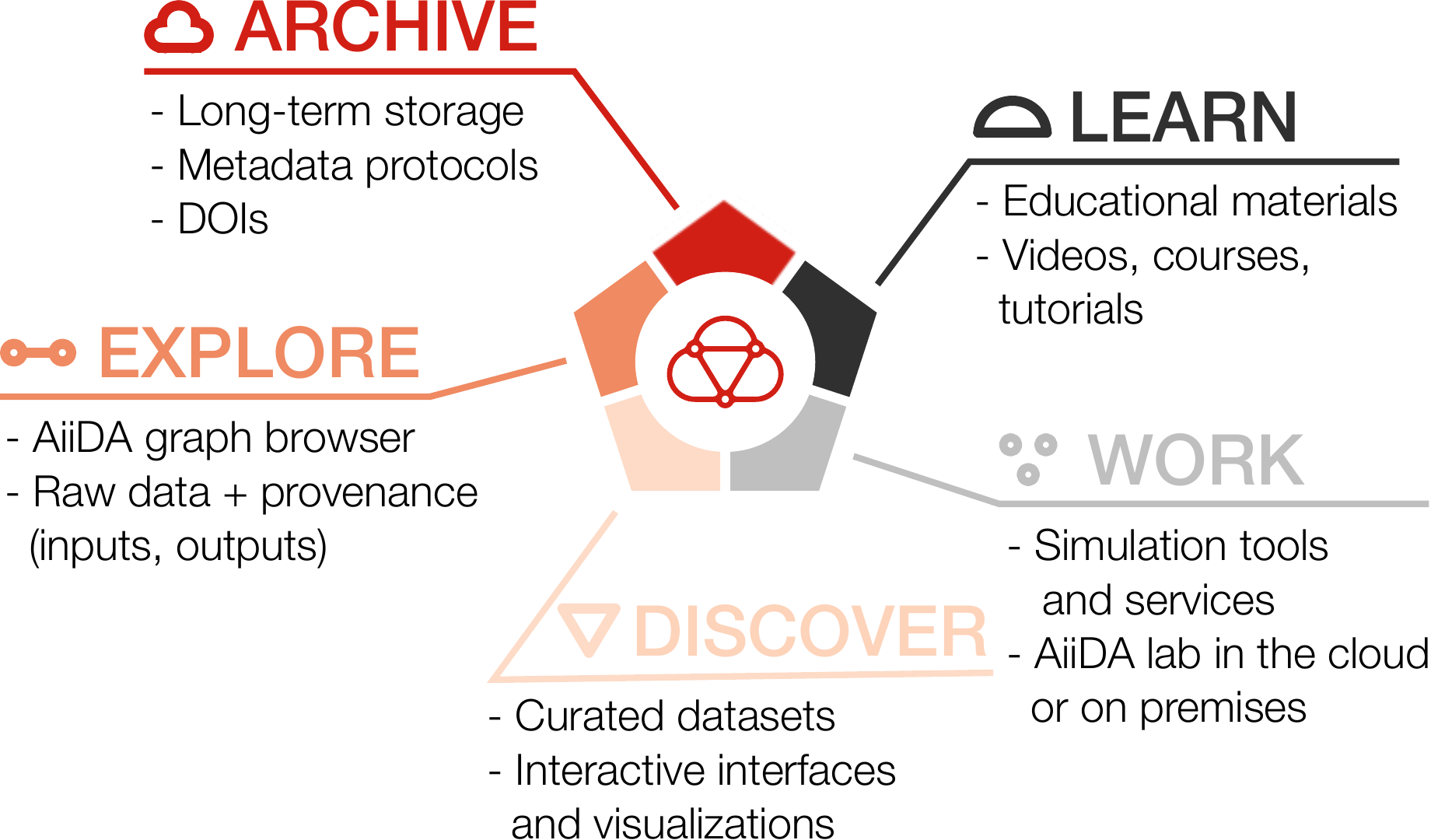}
\caption{\label{fig:overview}Materials Cloud organises its resources in five sections,
    \LEARN{}, \WORK{}, \DISCOVER{}, \EXPLORE{}, and \ARCHIVE{}, representing
    different stages of the research life cycle.
}
\end{figure}

Materials Cloud with its five sections -- \LEARN, \WORK, \DISCOVER, \EXPLORE{}, and \ARCHIVE{} -- aims to provide an ecosystem that supports researchers throughout the life cycle of a scientific project, and helps them make their research output FAIR and reproducible.
Fig.~\ref{fig:overview} illustrates how the five sections of Materials Cloud mirror the typical research cycle, from learning to
simulating and finally publishing curated results, which become the starting point for new research:
\LEARN{} (described in section \nameref{chap:outreach}) contains educational materials and videos;
\WORK{} (section \nameref{chap:turnkeysolutions}) focuses on simulation services, turnkey solutions and data analytics tools.
The three sections \DISCOVER{}, \EXPLORE{}, and \ARCHIVE{} are
Materials Cloud's approach to FAIR sharing of research data
(sections \nameref{chap:archive} and \nameref{chap:reproducibility}). 

Materials Cloud is powered by AiiDA, a workflow manager for computational science with a strong focus on provenance, performance and extensibility \cite{Pizzi2016a, AiiDA1}.
AiiDA plays two roles in this context: that of a manager of simulations, and that of a ``stenographer'' of events.
The manager lets scientists interact seamlessly with any number of remote high-performance computing (HPC) resources, and orchestrates computational workflows involving many steps, codes, and possible paths.
The stenographer records the data trail leading from the inputs to the results of a workflow, the \emph{data provenance}, and stores it in databases tailored for efficient data mining of heterogeneous results. 
Any such database can then be uploaded to the Materials Cloud, e.g., accompanying the submission of a scientific article, providing a comprehensive record of the research project.

While trying to ingest all results into one monolithic database provides advantages in terms of interoperability and data mining, it involves defining a schema which all future contributions need to fit into and adapt.
Materials Cloud avoids this limitation by adopting the ``repository of repositories'' model of GitHub et al., providing each submission with its own space.
By using the AiiDA provenance model, Materials Cloud contributors nevertheless benefit from a unified user experience for browsing and searching for data and simulations.
They can rely on standardised AiiDA data types, where appropriate, while AiiDA's flexible plug-in system allows to add new types or to extend existing ones to fit the specific purpose of the research undertaken.

Specifically, the \ARCHIVE{} is a moderated repository, where researchers can submit relevant research data from computational materials science in formats of their choice, including (but not limited to) AiiDA provenance graphs.
The repository guarantees long-term storage of records and associated metadata, their findability via persistent identifiers, and their accessibility via standard protocols. 
The \ARCHIVE{} can also form the basis for additional, interlinked layers of accessibility, interoperability and reusability: 
\DISCOVER{} allows researchers to adds curated visualisations for their data, providing intuitive interfaces and context,
while \EXPLORE{} provides access to the underlying raw and complete AiiDA provenance via an interactive graph browser.
In this model, AiiDA plays a role similar to Git (by tracking materials science simulations) while Materials Cloud plays the role of GitHub (a platform to share, browse and visualise all that has been tracked by AiiDA).

In the following, we present the individual sections of Materials Cloud in detail.

\subsection*[FAIR data]{FAIR data: \ARCHIVE{} and \DISCOVER{}} \label{chap:archive}

The \ARCHIVE{} and \DISCOVER{} sections allow researchers to make their data available in a 
findable, accessible, interoperable, and reusable (FAIR) way \cite{Wilkinson2016}. 
The Materials Cloud \ARCHIVE{} is an open-access, moderated repository for research data in computational materials science that allows researchers worldwide to upload and publish their data free of charge.
In particular:
\begin{itemize}
    \item it provides globally unique and persistent digital object identifiers (DOIs) for every record;
    \item metadata are always publicly available (Creative Commons Attribution Share-Alike 4.0 license);
    \item metadata can be harvested in a number of machine-readable formats,
        including HTML meta tags (Dublin core), OAI-PMH (Dublin core) and JSON-LD (schema.org);
    \item all data are stored at the Swiss National Supercomputing Centre;
    \item it is non-commercial and free of charge;
    \item data records are guaranteed to be preserved for at least 10 years after deposition;
    \item current size limits are 5 GB for general data records and 50 GB for AiiDA databases; 
    \item moderators can approve larger data sets upon request (currently, ~0.5 petabytes are allocated overall, with a 10-year retention time per record).
\end{itemize}

Data management plans (DMPs) that describe the handling of data both during a research project and after its completion are becoming standard components of applications for research grants.
The Materials Cloud \ARCHIVE{} is listed on the re3data\cite{re3data} and FAIR sharing\cite{FAIRsharingTeam2018} repository registries, indexed by Google Dataset Search and B2FIND (\href{http://b2find.eudat.eu/}{b2find.eudat.eu}), and it is a recommended repository for materials science by Nature Scientific Data.\cite{scidata}
It complies with the data repository requirements of major funding agencies, and provides tailored DMP templates (\href{https://www.materialscloud.org/dmp}{materialscloud.org/dmp}).

Unlike \emph{interdisciplinary} repositories for research data, such as Zenodo (\href{https://zenodo.org}{zenodo.org}), Data Dryad (\href{https://datadryad.org/}{datadryad.org}), the Open Science Framework (\href{https://osf.io/}{osf.io}), or figshare (\href{https://figshare.com/}{figshare.com}), the \ARCHIVE{} is \emph{moderated} and focuses on providing added value for datasets from computational materials science.
Submissions to the \ARCHIVE{} are expected to provide data that is of value to and can be used by other researchers in the field, such as data supporting a past, present or future peer-reviewed paper.
Materials Cloud moderators are subject experts, who follow a set of criteria (\href{https://www.materialscloud.org/moderation}{materialscloud.org/moderation}) to flag unsuitable or duplicate content, inappropriate form or topic, or excessive submission rates, much in the spirit of the arXiv preprint server (\href{https://arxiv.org}{arxiv.org}).
While all data formats are accepted, moderators will suggest alternative formats, where applicable, that improve interoperability and reusability, in line with the 5-star deployment scheme to open web data (\href{http://5stardata.info/en}{5stardata.info}).

\begin{figure}[!h]
\centering
\includegraphics[width=\textwidth]{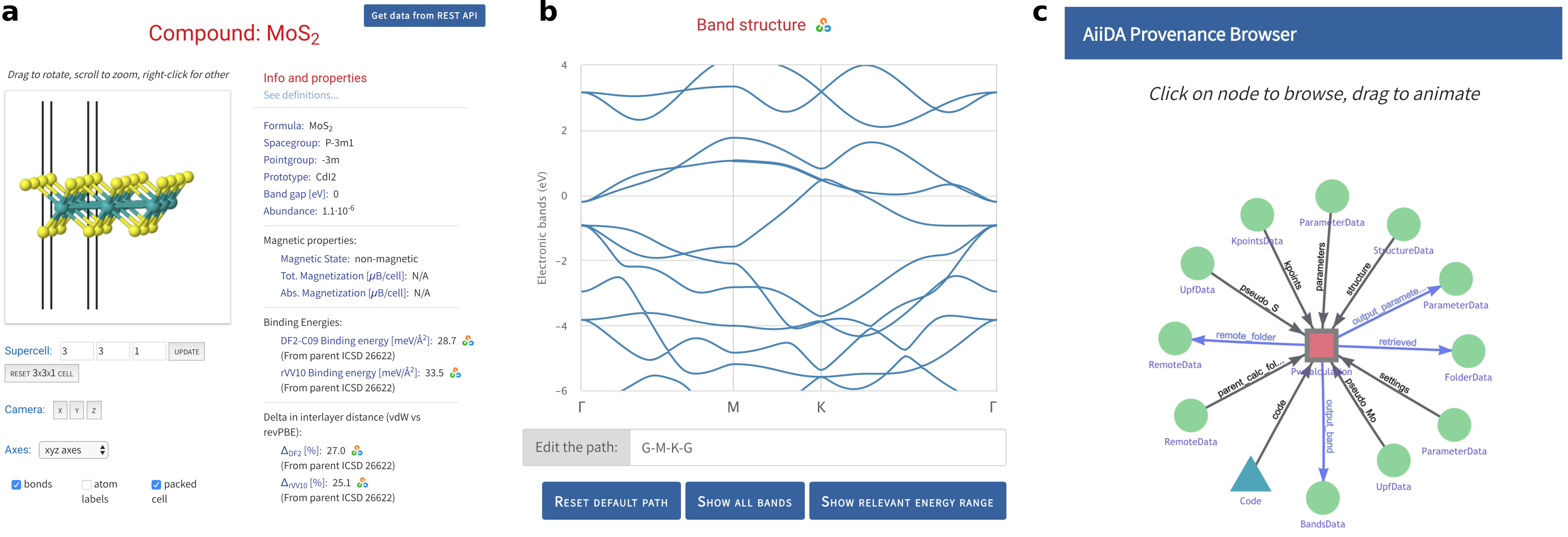}
\caption{\DISCOVER{} section on ``\emph{2D structures and
    layered materials}'' \cite{MounetArchive2018}. 
The ``ID card'' of a material displays key computed properties, as well as interactive visualisations of the crystal structure~\textbf{(a)}, the electronic band structure~\textbf{(b)} and more.
AiiDA icons link every piece of data to its corresponding node in the provenance graph that can be browsed through the \EXPLORE{} interface~\textbf{(c)}.
}

\label{fig:2Dmaterials}
\end{figure}

\begin{figure}[!h]
    \centering
    \includegraphics[width=\textwidth]{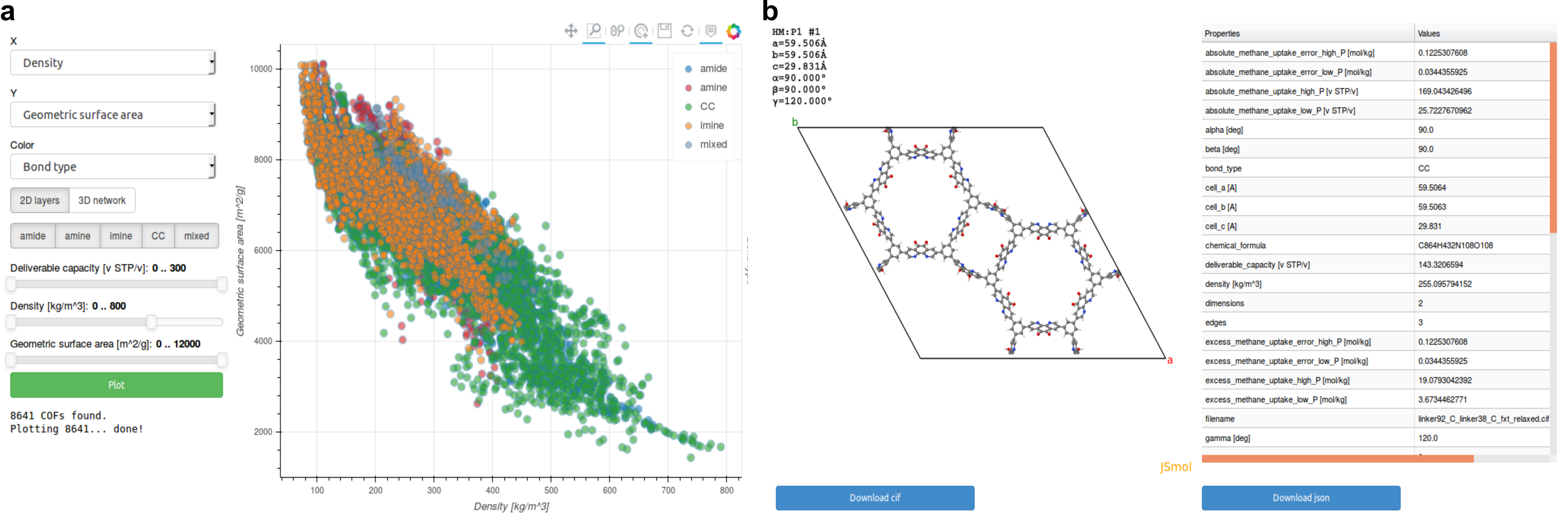}
    \caption{\DISCOVER{} section on ``\emph{Covalent organic frameworks (COFs)
        for methane storage applications}'' \cite{RocioArchive2018}, presenting almost 70000 COFs assembled
\emph{in silico}, together with their computed properties~\textbf{(a)} and
atomic structures~\textbf{(b)} in the form of interactive figures that mirror
those published in the corresponding peer-reviewed paper.}
\label{fig:sssp}
\end{figure}

Researchers can leverage the full power of the approach by adding interactive \DISCOVER{} and \EXPLORE{} interfaces to their datasets in order to provide further layers of accessibility, interoperability and reproducibility (see also section \nameref{chap:reproducibility} below).
\DISCOVER{} sections focus on curated data, presented in the form of dedicated interactive visualisations. 
For example, in the \DISCOVER{} section ``\emph{2D structures and layered materials}'' \cite{MounetArchive2018}, users can browse the curated dataset discussed in Ref.~\cite{Mounet2018}.
After selecting a material, key properties of the compound are displayed on a detail page (Fig.~\ref{fig:2Dmaterials}a,b), which includes interactive visualisations of quantities, such as the crystal structure, the electronic band structure, as well as phonon eigenvectors and band structures.
Fig.~\ref{fig:sssp} shows screenshots from another \DISCOVER{} sections on ``\emph{Covalent organic frameworks (COFs) for methane storage applications}''~\cite{RocioArchive2018}, containing interactive versions of the static figures published in reference \cite{Rocio2018}. 
The research data underlying all \DISCOVER{} sections on Materials Cloud is published in corresponding \ARCHIVE{} records and citable through DOIs.

What differentiates Materials Cloud \DISCOVER{} sections from other approaches to presenting materials data, is that each piece of data in a \DISCOVER{} section can be linked to a node in the AiiDA provenance graph (Fig.~\ref{fig:2Dmaterials}c) for full reproducibility, as we discuss in the following.

\subsection*[Reproducibility]{Reproducibility beyond FAIR: \EXPLORE{} }  \label{chap:reproducibility}

While making data FAIR simplifies and accelerates the sharing of knowledge, it is equally important to ensure that the knowledge being shared is reliable.
Computational materials science involves running computer programs on digital inputs and producing digital outputs. 
Yet, historically, only \emph{some} input and output data have been shared in the computational materials science literature, often in narrative form, making it unnecessarily difficult for peers to reproduce reported results. 
While storing and sharing \emph{all} data may not be technically feasible or financially sensible, researchers (and reviewers) today should demand that the data provided is sufficient to reproduce the reported results in their entirety.

This simple and seemingly self-evident demand can be tedious and time-consuming to meet in practice.
Researchers leave out pieces of information for a variety of reasons: data may appear trivial, irrelevant or too complex to provide in accessible form.
The challenge of providing access to this data is amplified, e.g. in studies involving large numbers of materials or workflows with many different steps, and calls for tools that simplify and automate this task.

\begin{figure}[!h]
    \centering
    \includegraphics[width=\textwidth]{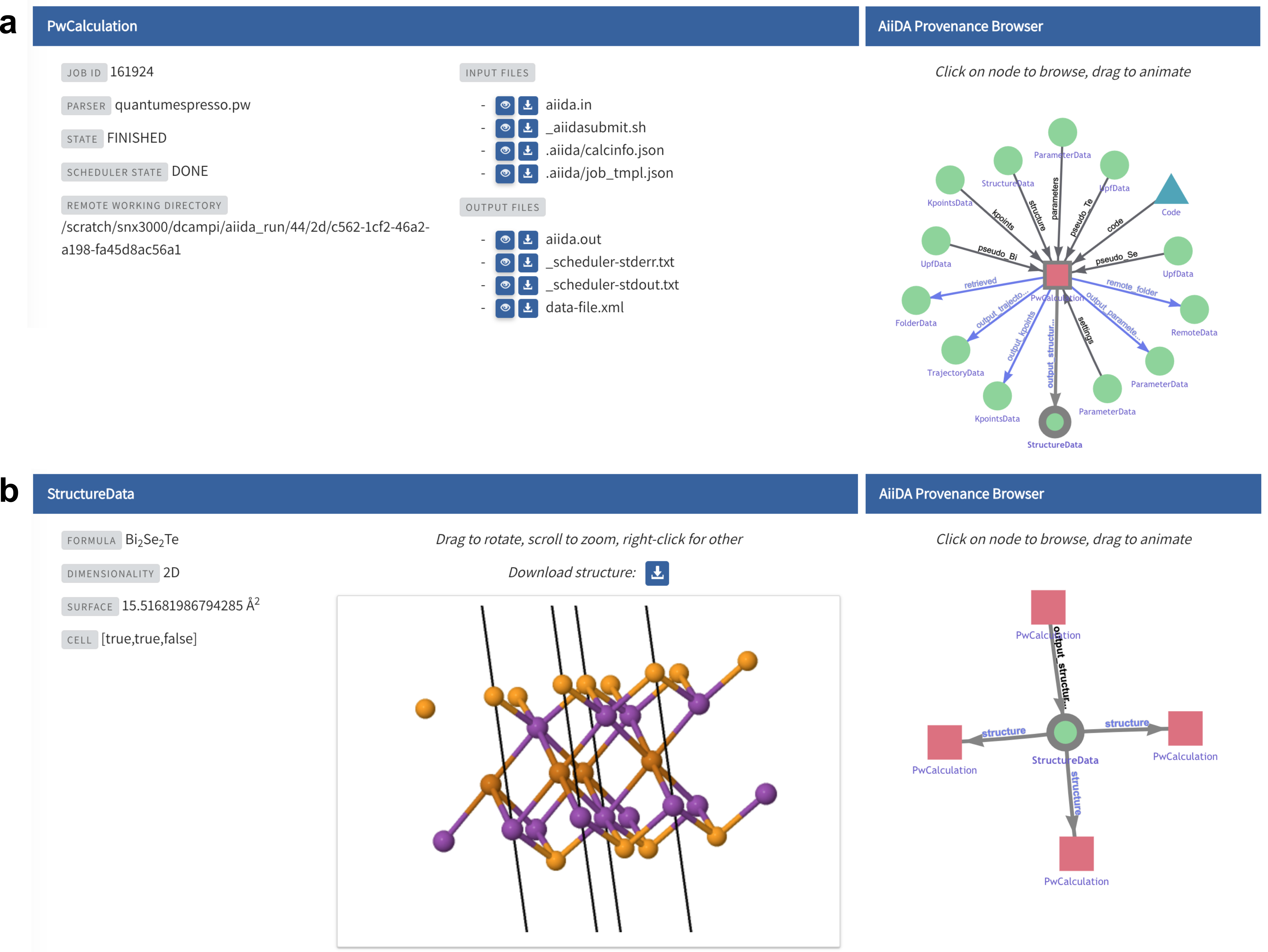}
    % Graph generated using: 
    % verdi -p mcloud_paper node graph generate 442dc -a 3 -d 3
    % => 54.dot.pdf
\caption{\EXPLORE{} interface for AiiDA provenance graphs.
    \textbf{(a)} Interactive view of a calculation node (here representing a run of \texttt{pw.x} code from the Quantum ESPRESSO suite~\cite{Giannozzi2017}), providing download links for all input and output files.
    The provenance browser on the right allows to jump to the visualisation of any input or output node of the calculation.
    \textbf{(b)} Interactive view of the atomic crystal structure returned by calculation \textbf{(a)}. The provenance browser indicates this structure was used in three subsequent calculations.
    See the supporting information for the full provenance graph.
}
\label{fig:provenance}
\end{figure}

In AiiDA, the ``stenographer'' records, for every calculation, a set of standardised data and metadata in a dedicated database \cite{aiida-orm}.
This includes information on who submitted the calculation, when the calculation was submitted, which computer and code were used, which inputs were used, which outputs were produced, as well as how these outputs are further used as inputs to the next calculation (see also Fig.~\ref{fig:provenance}).
Since long-term data storage is more expensive than the short-term storage used by active simulations, it is often not reasonable to preserve all output data.
Which output data is stored is decided by the AiiDA plug-in for the code in question -- for example, in a density-functional theory calculation, total energies, electronic band structures and log files might be stored by default, while Kohn-Sham wave functions might be discarded.
The overarching principle, however, is that \emph{all information needed to reproduce the outputs must be preserved}, even if not all intermediate files are persisted.
By combining this information stored at the level of individual calculations with the logical relationships between successive calculations, AiiDA provides reproducibility of entire workflows out of the box. 

Scientists who use AiiDA for their calculations can choose to upload their AiiDA databases to the \EXPLORE{} section in order to complement their published research with a complete record of their calculations.
When they do so, peers can browse the AiiDA graph as shown in Fig.~\ref{fig:provenance}:
all nodes of the graph, representing either calculations or pieces of data, can be inspected and are linked to their parent and children nodes via the provenance browser.
Dedicated visualisations make the content of nodes intuitively accessible, and allow to download individual pieces of data  (such as crystal structures or input files).
Subject experts, on the other hand, can install AiiDA on their computer, import the AiiDA database and continue their own research from where the authors of the original work left off.

We note that the interactive provenance browser is not limited to datasets uploaded to Materials Cloud:
AiiDA users can connect their own database to the \EXPLORE{} JavaScript application (via AiiDA's built-in REST API) and directly browse their own database without their data ever leaving their computer.

\subsection*[Simulation services]{Simulation services: \WORK{} and \AIIDALAB{}} \label{chap:turnkeysolutions}

While \DISCOVER{}, \EXPLORE{}, and \ARCHIVE{} enable the dissemination of results that have already been computed, \WORK{} aims to facilitate data generation and analytics by means of simulation tools and services.
The \WORK{} section leverages web technologies in order to make well-defined calculations and workflows simple to run and accessible to a wide user base, including students, experimental scientists, and computational scientists.

\begin{figure}[!h]
\centering
\includegraphics[width=1\textwidth]{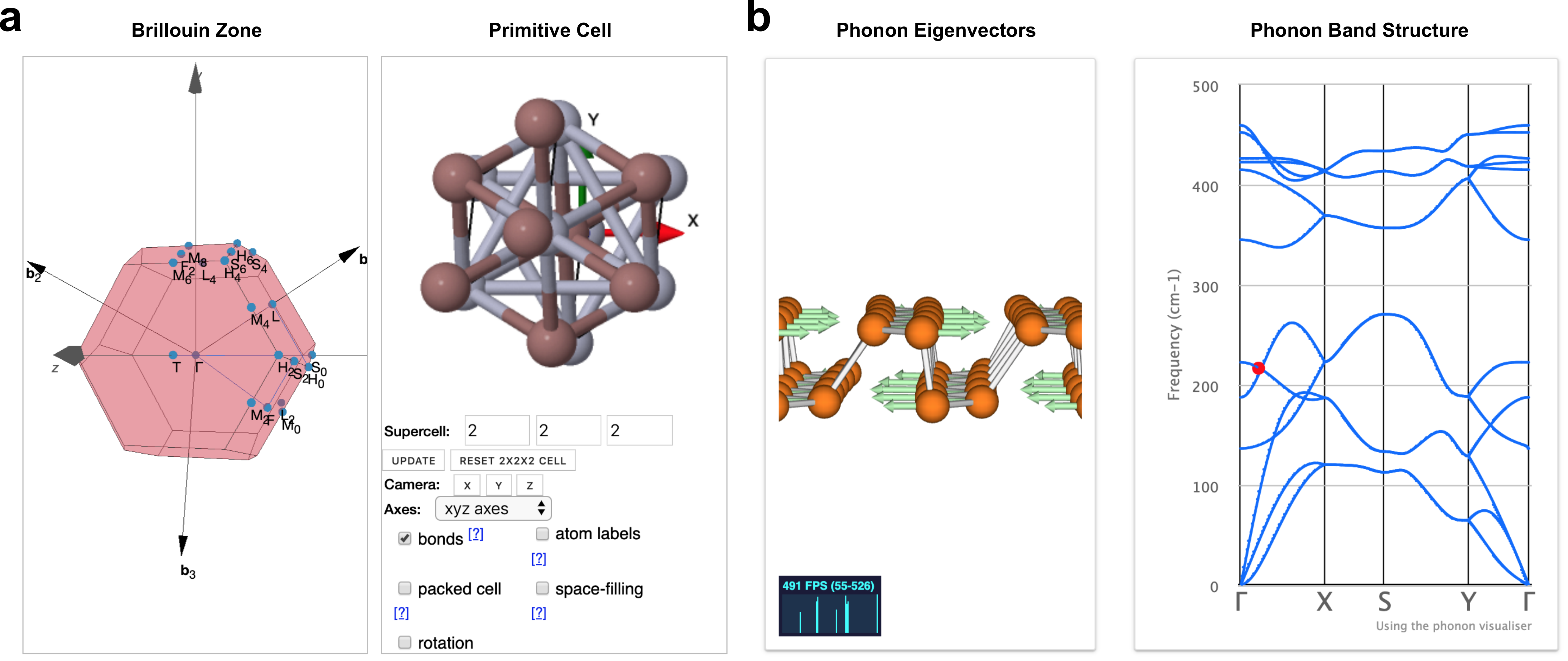}
\caption{Tools in the \WORK{} section. 
    \textbf{(a)} SeeK-path tool \cite{Hinuma2017} for finding and visualising paths in reciprocal space, here showing the Brillouin zone of InHg.
    \textbf{(b)} Interactive visualiser for lattice vibrations
(adapted from \href{http://henriquemiranda.github.io/phononwebsite/}{henriquemiranda.github.io/phononwebsite}), here for two-dimensional phosphorene.
Shown is the phonon eigenvector (left) corresponding to the red dot in the phonon band structure (right).
}
\label{fig:mc-tools}
\end{figure}

On the one hand, this includes stand-alone tools that run computationally inexpensive simulations, which produce immediate results:
for example, tools that help with plotting electronic band structures (Fig.~\ref{fig:mc-tools}a) or visualising lattice vibrations (Fig.~\ref{fig:mc-tools}b), and several tools leveraging machine learning methods.
The underlying docker technology (\href{https://www.docker.com}{docker.com}) makes it possible to support a diverse set of software frameworks on the same platform, allowing for custom solutions that are adapted to the specific tool in question.

\begin{figure}[!h]
\centering
\includegraphics[width=1\textwidth]{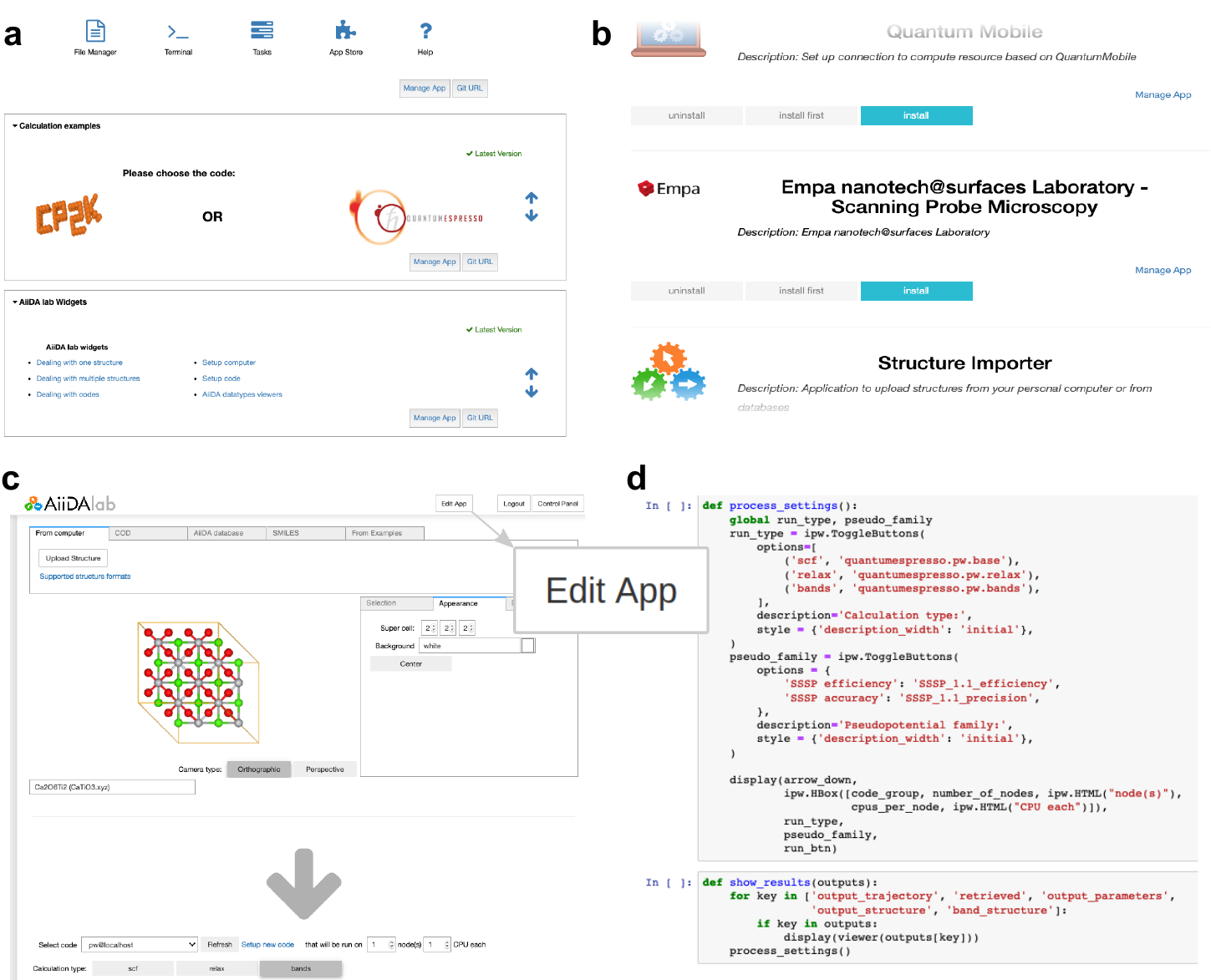}
\caption{\AIIDALAB{} simulation environment. 
    \textbf{(a)} Landing page with an overview of the applications installed. 
    \textbf{(b)} ``App store'' for managing applications.
    \textbf{(c)} Application that computes the optimised crystal structure of an input material as well as its electronic band structure along standardised paths.
    Clicking ``Edit App'' switches to the source code editor \textbf{(d)} of the underlying Jupyter notebook.
}
\label{fig:aiidalab-app}
\end{figure}

On the other hand, the \WORK{} section focuses on the \AIIDALAB{}, an ecosystem for applications powered by the AiiDA workflow manager (\href{https://materialscloud.org/aiidalab}{materialscloud.org/aiidalab}).
\AIIDALAB{} aims to remove barriers related to the set up and installation of simulation software by providing access to applications for launching and controlling computational workflows directly from the web browser.
Users log on to a private, containerised environment that provides a persistent work space for storing apps, the AiiDA database, and file repositories (Fig.~\ref{fig:aiidalab-app}a,b).
\AIIDALAB{} apps let users connect to their own computational resources anywhere in the world in order to run production-grade workflows.
Users can import data into the platform either by uploading from their computer, or by importing data from connected open databases such as the Crystallography Open Database \cite{Grazulis2012} or any database implementing the OPTIMADE standard (\href{https://www.optimade.org/}{optimade.org}), including AFLOW (\href{http://aflow.org/}{aflow.org} \cite{Curtarolo2012}), COD (\href{http://www.crystallography.net/cod/}{crystallography.net/cod} \cite{Grazulis2012}), TCOD (\href{http://www.crystallography.net/tcod/}{crystallography.net/tcod} \cite{Grazulis2014a}), Materials Cloud, MPDS (\href{https://mpds.io}{mpds.io} \cite{Blokhin2018}), Materials Project (\href{https://materialsproject.org/}{materialsproject.org} \cite{Jain2013}), NOMAD (\href{https://nomad-coe.eu/}{nomad-coe.eu} \cite{Ghiringhelli2017}), Open Materials Database (\href{http://openmaterialsdb.se/}{openmaterialsdb.se}\cite{httk}), and OQMD (\href{http://oqmd.org/}{oqmd.org} \cite{Saal2013}).

The intuitive graphical interface makes \AIIDALAB{} applications an ideal vehicle for sharing turnkey solutions with non-specialists, be it computational scientists from another discipline or experimental researchers with no programming experience.

% Trick to get rid of whitespace on the right of jupyter notebook layouts:
% Untick 'max-width: calc(100% - 14ex);' on output_subarea div

From a technological perspective, \AIIDALAB{} applications are Jupyter notebooks (\href{http://jupyter.org}{jupyter.org}) containing instructions for the AiiDA workflow manager, which are transformed into an interactive web application (see Fig.~\ref{fig:aiidalab-app}c,d).
This design has two important implications for app \emph{development}: 
First, the widespread adoption of Python and Jupyter notebooks in data science
in general~\cite{Perkel2018} and computational materials science in particular
makes most researchers in the field potential app developers.
In particular, thanks to Jupyter widgets, interactive web interfaces can be written in a few lines of Python, and no longer require knowledge of JavaScript).
And second, AppMode lets developers switch between the graphical app layout (Fig.~\ref{fig:aiidalab-app}c) and the Python development environment (Fig.~\ref{fig:aiidalab-app}d) at the click of a button.
Apps can be edited live in the browser, and developers have the full power of the Python programming language at their fingertips.

\AIIDALAB{} encourages sharing of workflows and visualisations via an App store model: in a first step, developers register their application on the application registry (\href{https://aiidalab.github.io/aiidalab-registry/}{aiidalab.github.io/aiidalab-registry}).
Once registered, users can then install the app via the built-in application manager (Fig.~\ref{fig:aiidalab-app}b) and access it from their home screen (Fig.~\ref{fig:aiidalab-app}a).
The source codes of the \AIIDALAB{}, AppMode, and AiiDA itself are released under the permissive MIT open-source license (see code availability statement), enabling re-deployment of the \AIIDALAB{} platform both in academic and in corporate environments.

When a local installation is desired, e.g., for educational purposes, users can download the Quantum Mobile virtual machine (see section \nameref{chap:outreach} for a full description), which provides the same environment in a self-contained form and runs on Linux, MacOS, and Windows.

\subsection*[Education and outreach]{Education and outreach: \LEARN{} and \QM{}} \label{chap:outreach}

\begin{figure}[tbp]
    \centering
    \includegraphics[width=\textwidth]{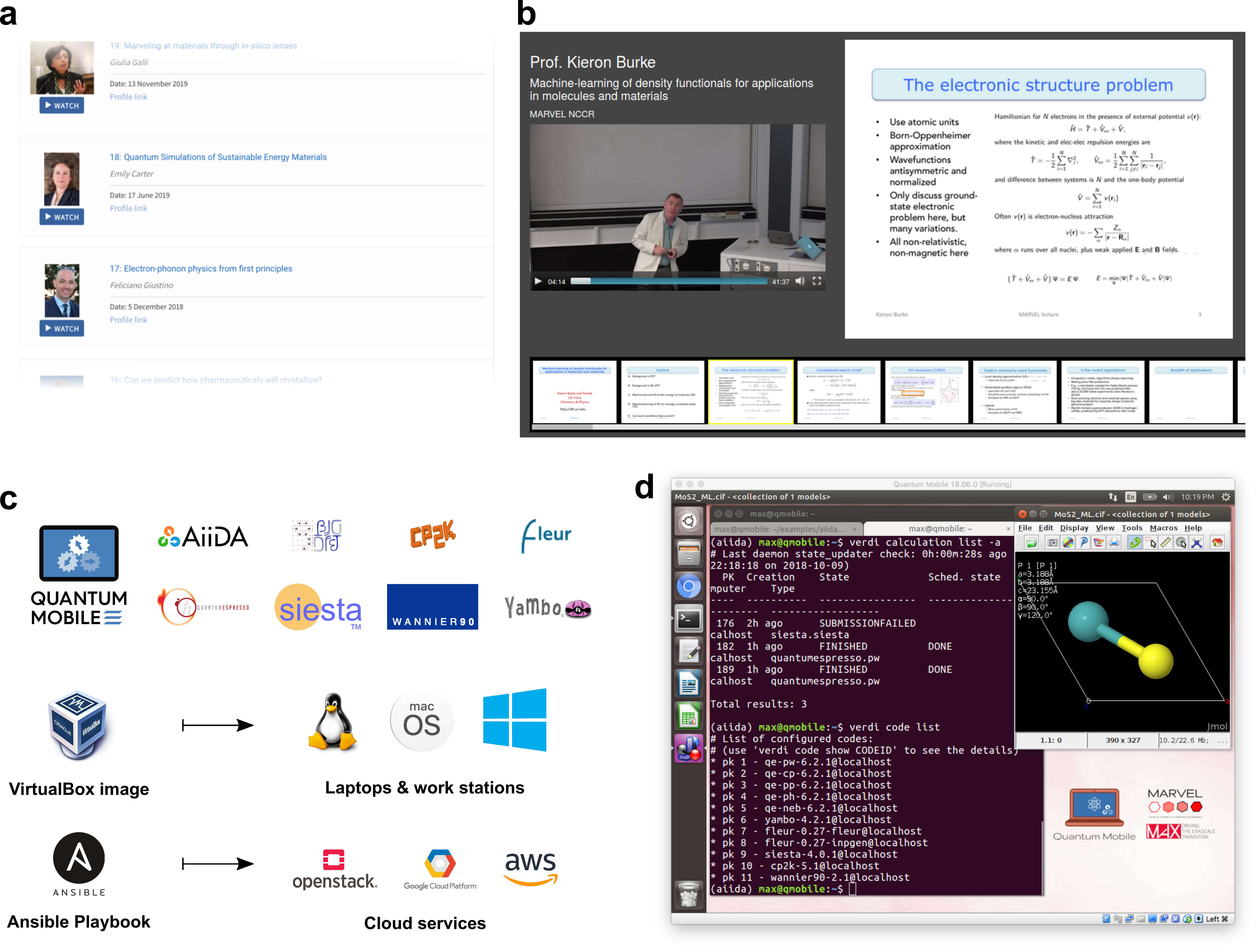}
\caption{\label{fig:outreach}Education and outreach.
    (a) MARVEL distinguished lectures available in the \LEARN{} section.
    (b) Slideshot player with slide synchronisation and slide-based browsing.
    (c) Simulation codes provided with the Quantum Mobile virtual machine, and deployment schemes.
    (d) Screenshot of the Quantum Mobile desktop.
}
\end{figure}

The \LEARN{} section of Materials Cloud hosts video lectures, tutorials, and seminars in computational materials science (Fig. \ref{fig:outreach}a).
Lectures in collaboration with CECAM (\href{https://www.cecam.org/}{cecam.org}) include the ``Classics on Molecular and Materials Simulations'', dedicated to record pioneering contributions in the field, and the ``Mary Ann Mansigh conversations'' in which outstanding representatives from computational science share their perspective on how modelling affects society.
Videos are grouped by topic or event, presented together with accompanying materials, and slides where available. 
The Slideshot video player shows video and slides side by side, and keeps them in sync (Fig. \ref{fig:outreach}b, \href{https://slideshot.epfl.ch/}{slideshot.epfl.ch}). 

Besides the educational materials in the \LEARN{} section, students can also download the Quantum Mobile virtual machine for computational materials science (Fig. \ref{fig:outreach}c) from the \WORK{} section.
Quantum Mobile is based on Ubuntu Linux and comes pre-installed with a collection of open-source software packages for quantum-mechanical calculations including
    Quantum ESPRESSO~\cite{Giannozzi2017},
    Yambo~\cite{Marini2009},
    fleur~\cite{fleur},
    Siesta~\cite{Soler2002},
    CP2K~\cite{Hutter2014},
    and Wannier90~\cite{Mostofi2008}. 

    Furthermore, it includes the Standard Solid State Pseudopotential Library (SSSP)~\cite{SSSPpaper,PrandiniArchive2018}, various visualisation tools (jmol~\cite{jmol}, XCrySDen~\cite{Kokalj1999}, gnuplot, grace), a job scheduler (Slurm) and a build environment with C, C++ and Fortran compilers as well as scientific and MPI libraries.
AiiDA and the \AIIDALAB{} environment are pre-configured, including AiiDA plug-ins for each of the ab initio codes listed above, ready to be used out-of-the-box (Fig.~\ref{fig:outreach}c,d).

Quantum Mobile provides a uniform environment for quantum mechanical materials simulations and runs on most popular operating systems, including Linux, MacOS and Windows, via the VirtualBox software (\href{https://www.virtualbox.org}{virtualbox.org}).
Contrary to other encapsulation strategies, such as Docker, students interact with a familiar graphical desktop, shown in Fig.~\ref{fig:outreach}d.
Since its first release in November 2017, Quantum Mobile has been continuously updated and was used in lecture courses at EPFL, ETHZ, and Ghent University (\href{http://compmatphys.org}{compmatphys.org}) as well as in numerous tutorials on electronic structure methods, molecular simulations, and AiiDA (see \href{https://www.materialscloud.org/quantum-mobile}{materialscloud.org/quantum-mobile}), where it helps to reduce the time needed for installation and configuration of software. 

The modular design of Quantum Mobile takes into account that one size does not fit all:
its components (simulation codes, tools, data) are encapsulated in reusable, individually tested components (see Code availability statement).
Teachers can pick and choose from a growing repository of more than 30 roles and build their own version of Quantum Mobile containing just the tools they need.

\label{chap:technology}

\section*{Discussion and Outlook}

The increasing availability and standardisation of infrastructure-as-a-service (IaaS) make it possible to share the  findings and capabilities developed by computational materials science not only with peers who possess journal subscriptions and specialist software, but with anyone familiar with using a web browser.
In the case of Materials Cloud, this includes (i) the interconnected outcomes of calculations and workflows, presented in a findable, accessible, interoperable, reusable, and reproducible way (\DISCOVER{}, \EXPLORE{}, and \ARCHIVE{} sections), as well as (ii) turnkey solutions that launch state-of-the-art workflows at the click of a button (\WORK{} section).

Materials Cloud and AiiDA form the core of the open science platform used at the National Centre on Computational Design and Discovery of Novel Materials (MARVEL NCCR, started in 2014), funded by the Swiss National Science Foundation,
the Centre of Excellence for Materials Design at the Exascale (MaX, started in 2015), funded by the European Commission, as well as further partner projects (\href{https://www.materialscloud.org/home\#partners}{materialscloud.org/home\#partners}).
Since its official launch in early 2018, Materials Cloud has grown steadily as it becomes the central repository for sharing research data, workflows, and tools within MARVEL, MaX, and further partner projects.
Today, the Materials Cloud \ARCHIVE{} provides a moderated repository for the long-term storage of materials science research data, open to submissions from around the world.
For \AIIDALAB{}, we propose a model where interested parties, such as academic institutes, research centres, and companies can re-deploy the open-source platform on their own (virtual) hardware.

While the content on Materials Cloud can indeed be \emph{used} simply through a web browser, \emph{submitting} new tools and interactive visualisations still requires technical expertise. 
We are working both on lowering this barrier and on reducing the associated workload from platform administrators by moving in the direction of a platform-as-a-service architecture.
The submission procedure and interface of the \ARCHIVE{} will soon be further streamlined by the switch to the Invenio framework (\href{https://invenio-software.org}{invenio-software.org}), bringing user authentication, search and more.
Finally, the governance model of Materials Cloud will evolve, adapting to its increasing role within the MARVEL and MaX scientific centres, and the computational materials science community at large.

One unresolved challenge in the field is the task to find a common language for information exchange between OSPs.
Efforts to move forward in this direction range from collecting existing semantic assets in computational materials science \cite{VasilyBunakov2018}, over the design of new ontologies \cite{emmo-gh}, to specifications of interoperable data formats \cite{escdf,nomad-metainfo}, and application programming interfaces  (\href{https://www.optimade.org/}{optimade.org}).
Once these efforts converge, they can be connected to existing infrastructures for structured web data (\href{https://schema.org}{schema.org}).

Another important challenge is to secure long-term support for continued development.
The diversity of relevant services goes far beyond the long-term storage of files and requires maintenance.
Analogies can be drawn to other major research infrastructures, ranging from particle accelerators over telescopes to libraries, where key services are provided to the scientific community, either by the public or in the form of public-private partnerships.
Given the unprecedented availability of computational power (\href{https://www.top500.org/}{top500.org}), 
the pervasiveness of computational (materials) science in the scientific literature \cite{VanNoorden2014}, 
and its relevance to pressing societal challenges \cite{Satell2019}, 
maintaining functional research infrastructures for computational science -- at comparatively low cost -- would seem like a forward-looking investment.

\section*{Methods}

\begin{figure}[!h]
\centering
\includegraphics[width=\textwidth]{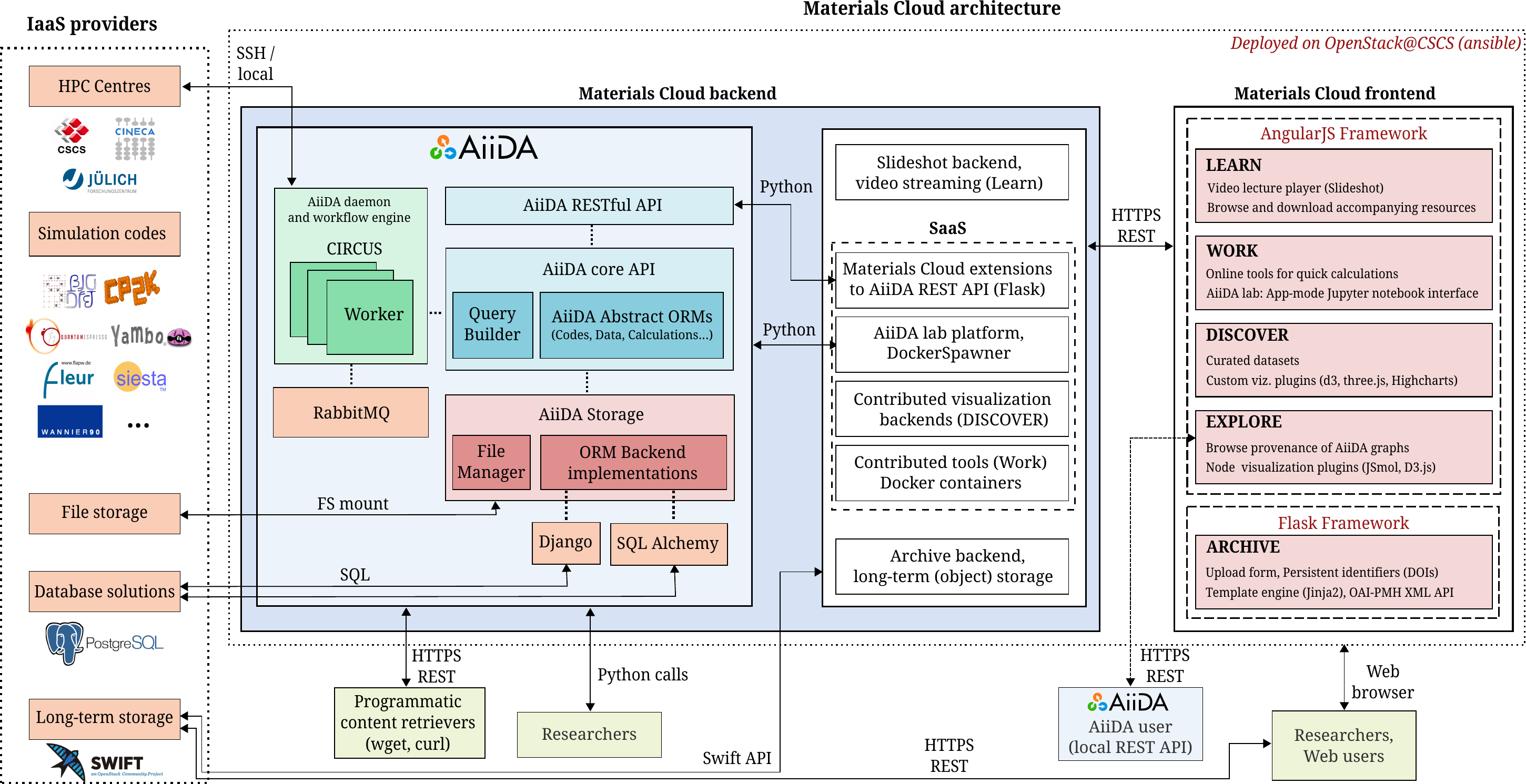}
\caption{\label{fig:architecture}Materials Cloud architecture diagram.
 Independent frontends for \LEARN{}, \WORK{}, \DISCOVER{}, and \EXPLORE{} based on 
 AngularJS are powered by different backends, including
 AiiDA REST APIs, tools encapsulated in docker containers and a JupyterHub running
 one docker container per AiiDA user.
}
\end{figure}
Materials Cloud's modular architecture, sketched in Fig.~\ref{fig:architecture}, is designed to enable updates of individual sections without affecting the rest of the service.
The top-level web user interface is presented through a set of AngularJS applications with one app per section.
Section content is served either directly by the corresponding application (\LEARN{}, \DISCOVER{}, \EXPLORE{}) or through user interfaces provided by containerised content (\WORK{}, \DISCOVER{}).
The overall Materials Cloud theme is based on the Bootstrap~(\href{https://getbootstrap.com}{getbootstrap.com}) and Material Design~(\href{https://material.angularjs.org/latest/}{material.angularjs.org}) JavaScript libraries;
and individual sections use a range of visualisation libraries, including Highstock/Highcharts~(\href{https://www.highcharts.com}{highcharts.com}), D3js~(\href{https://d3js.org}{d3js.org}), JSMol~(\href{http://jsmol.sourceforge.net/}{jsmol.sourceforge.net/}), and Vis~(\href{http://visjs.org}{visjs.org}).

A slideshot server provides the API to serve videos and slides to the \LEARN{} section. 
Tools in the \WORK{} section are encapsulated in docker containers and control their own web frontends.
The AiiDA lab is a customised JupyterHub that is isolated from the rest of the platform and runs on a separate server.
Every \AIIDALAB{} account is associated with a private container, including persistent storage and compute resources, and may be set up to connect to high-performance computing resources owned by the account holder.
Containerised contributions to \WORK{} and \DISCOVER{} may use different Python-based frameworks, such as Flask~(\href{https://flask.palletsprojects.com/}{flask.palletsprojects.com}), Django~(\href{https://www.djangoproject.com/}{djangoproject.com}), or Bokeh~(\href{https://bokeh.org/}{bokeh.org}).

In the \EXPLORE{} section, the frontend JavaScript application talks directly to the standardised AiiDA application programming interface (API).
This representational state transfer (REST) API provides access to calculations, workflows, codes, and data stored in the AiiDA graph, and makes them available in the JavaScript Object Notation (JSON) format.
The AiiDA REST API ships together with AiiDA, and besides serving static AiiDA databases on the Materials Cloud, AiiDA users can take advantage of the same JavaScript application to browse their own local AiiDA database.
For more details, see Fig.~S1 in the supplementary materials.

The \ARCHIVE{} section is only loosely coupled to the rest of the platform.
Files associated with \ARCHIVE{} records are stored in an OpenStack Swift Object Store and backed up to tape daily (\href{https://user.cscs.ch/storage/object_storage/#data-protection}{user.cscs.ch/storage/object\_storage}).
The \ARCHIVE{} server hosts the database containing the metadata associated with records, and delegates requests for associated files to the object store via short-lived unique URLs.
The current implementation is based on the Flask microframework, but will transition in 2020 to a highly scalable infrastructure based on Invenio 3, the open-source Python framework powering the Zenodo repository operated by CERN.

Materials Cloud is deployed on virtual machines running in an OpenStack cloud computing platform~(\href{https://www.openstack.org/}{openstack.org}) at the Swiss National Supercomputing Centre (CSCS).
All production servers are duplicated, following standard web development practises (see Fig.~S2 in the supplementary materials for details).
In order to prevent loss of log files and user data, backups are taken periodically and stored in the object storage service at CSCS. 
A server at a different physical location monitors availability and basic functionality of all production services every 60 seconds and notifies maintainers in case of unexpected deterioration of service.

Deployment is automated using Ansible playbooks~(\href{https://www.ansible.com}{ansible.com}), which allow software provisioning, configuration management, and application deployment on remote machines over SSH.
The use of automated Ansible roles, together with Materials Cloud's modular architecture and the widely available OpenStack infrastructure, simplifies the redeployment of Materials Cloud (or components of it) in other locations, e.g., for the purpose of load balancing, federation of service, or in-house use.

\subsection*{Data availability}

The datasets discussed in this manuscript \cite{MounetArchive2018,RocioArchive2018},
as well as the datasets underlying all Materials Cloud \DISCOVER{} and \EXPLORE{} sections are available on the Materials Cloud Archive (\href{https://archive.materialscloud.org}{archive.materialscloud.org}) under Creative Commons licenses.

\subsection*{Code availability}

The source code of AiiDA, the \AIIDALAB{}, Appmode (\href{https://github.com/oschuett/appmode}{github.com/oschuett/appmode}), 
and most \AIIDALAB{} applications is released under the MIT open-source license, and made available under the \texttt{aiidateam} (\href{https://github.com/aiidateam}{github.com/aiidateam}) and \texttt{aiidalab} (\href{https://github.com/aiidalab}{github.com/aiidalab}) GitHub organisations.

The Quantum Mobile virtual machine can be downloaded from \href{https://materialscloud.org/quantum-mobile}{materialscloud.org/quantum-mobile}. 
Its source code (in the form of ansible roles and playbooks) is released under the MIT license and made available under the \texttt{marvel-nccr} GitHub organisation (\href{https://github.com/marvel-nccr}{github.com/marvel-nccr}).

\section*{Acknowledgements}

This work is supported by the MARVEL National Centre for Competency in Research funded by the Swiss National Science Foundation (grant agreement ID~51NF40-182892), 
the European Centre of Excellence MaX ``Materials design at the Exascale'' (grant no.~824143), 
the ``MaGic'' project of the European Research Council (grant agreement ID~666983),
the swissuniversities P-5 ``Materials Cloud'' project (grant agreement ID~182-008),
the ``MARKETPLACE'' H2020 project (grant agreement ID~760173),
the ``INTERSECT'' H2020 project  (grant agreement ID~814487),
the ``NFFA'' H2020 project (grant agreement ID~654360), the ``EMMC'' H2020 project (grant agreement ID~723867).
We acknowledge PRACE for awarding us simulation time on Piz Daint at CSCS (project ID~2016153543) and Marconi at CINECA (project ID~2016163963), 
the Swiss Platform for Advanced Scientific Computing PASC for the SIRIUS co-design activities,
and EPFL and the Swiss National Science Foundation for supporting our long-term data storage needs.

We thank the IaaS support teams at the Swiss National
Supercomputing Centre (CSCS): 
Sadaf Alam,
Vincenzo Annaloro, 
Marco Consoli, 
Pablo Fernandez, 
Stefano Gorini, 
Hussein Harake, 
Guy-Maël Horclois Le Pironnec,
Mark Klein,
Giuseppe Lo Re, 
Colin McMurtrie 
and Marco Passerini.
And, finally, we would like to thank the early contributors to Materials Cloud for helping shape the project through their feedback and submission of tools, of \DISCOVER{} and of \EXPLORE{} sections:
Nicolas Mounet,
Antimo Marrazzo,
Nicolas Hörmann,
Gianluca Prandini,
Yoyo Hinuma,
Félix Musil,
David M. Wilkins,
Michele Ceriotti,
Bonan Zhu,
Henrique Miranda,
Thibault Sohier,
Mohammad Moosavi,
QuanSheng Wu,
Oleg Yazyev,
Benjamin Meyer,
Clémence Corminboeuf,
Kevin M. Jablonka,
Daniele Ongari,
and Jinhui Guo.

\section*{Author contributions}

GP and NM conceived the project.

FG, LT, MU, BS, GP, and NM designed the structure of the Materials Cloud.

LT, SK, FG, OS, MU, SPH, and GP designed the software and service architecture.

SK, LT, AVY, EP, OS, MB, VG, FG, and GP implemented and deployed the platform.

SK, EP, AVY, MB, LT, VG and GP form the Materials Cloud Team currently maintaining the platform.

SPH, SZ, CWA, CSA and MU supported the Materials Cloud through development of required features in the AiiDA framework.

BS, CAP, DP, LT, GP, and NM supervised the implementation of different aspects of the platform.

JV, LT, GP and TCS coordinated integration of the platform with infrastructure at CSCS.

All authors discussed the manuscript.

\section*{Competing interests}

The authors declare no competing interests.

\printbibliography
\clearpage

\appendix
\renewcommand\thefigure{S\arabic{figure}}
\section{Supporting Information}
\setcounter{figure}{0}

\begin{figure}[!ht]
\centering
\includegraphics[width=\textwidth]{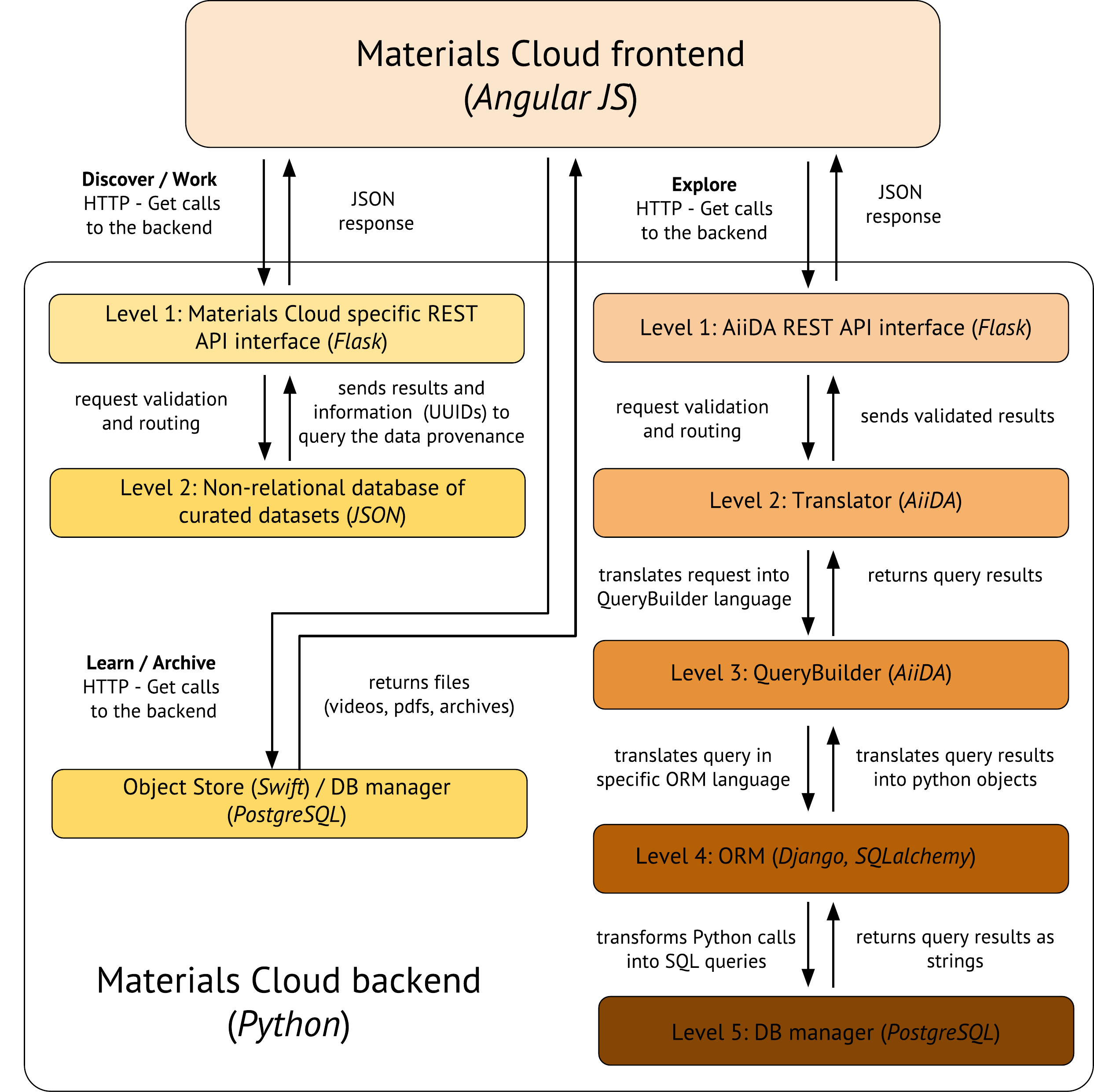}
\caption{\label{fig:rest-architecture}
    Data flow between Materials Cloud frontend and backend.
    Left: Data flow for \ARCHIVE{}, \DISCOVER{} and \WORK{}.
    Right: AiiDA REST API, flow starting from the browser request (top) via data validation, parsing and translation into the AiiDA query language down to the database query (bottom). 
    Response data then follows the reverse path and is returned to the browser in JSON format.
}
\end{figure}

\begin{figure}
\centering
\includegraphics[width=\textwidth]{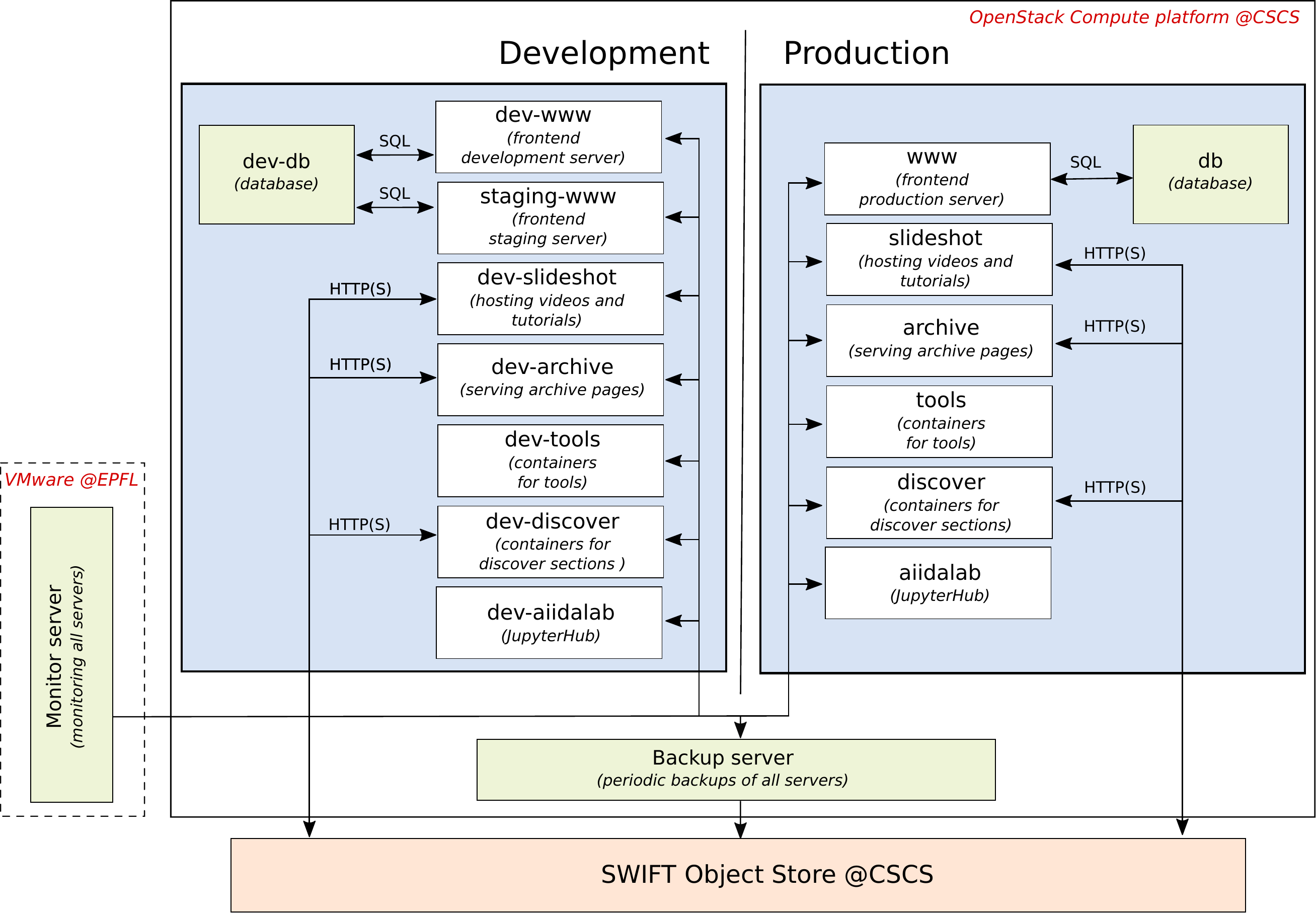}
\caption{Materials Cloud deployment diagram.
    Services are split across multiple virtual machines, 
    each with at least one clone for development and testing.
}
\label{fig:deployment}
\end{figure}

\begin{figure}
\centering
\includegraphics[width=\textwidth]{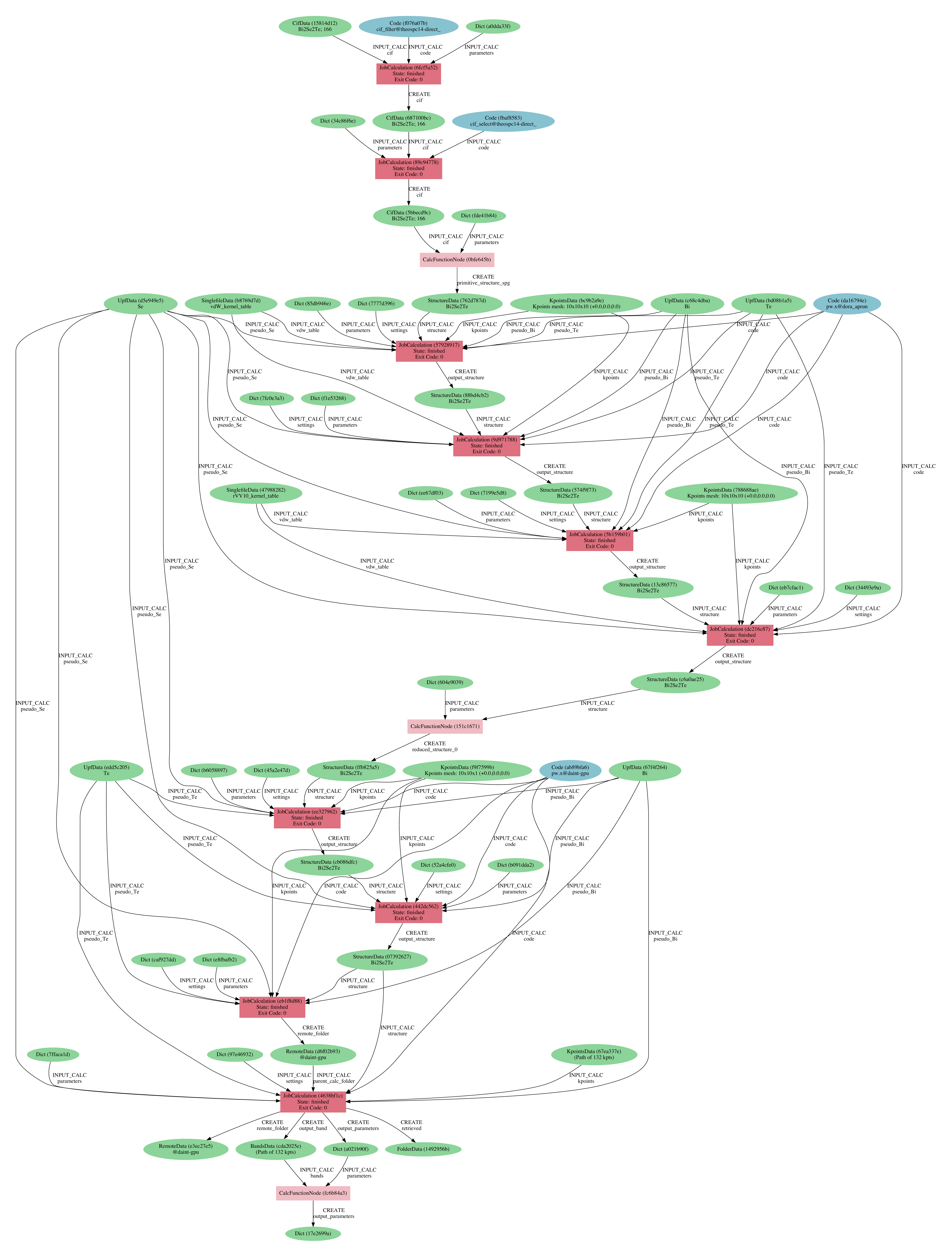}
\caption{Complete AiiDA provenance graph for the electronic band structure calculation of \ce{Bi2Se2Te}, as generated by \texttt{verdi node graph generate 4638bfc1} on the AiiDA database from reference \cite{MounetArchive2018}.
}

\label{fig:explore-graph}
\end{figure}

\end{document}